\documentclass[a4paper,fleqn,usenatbib]{mnras}
\usepackage{amsmath}
\usepackage{mathptmx}
\usepackage{txfonts}
\usepackage[T1]{fontenc}
\usepackage{ae,aecompl}
\usepackage{graphicx}	
\usepackage{amssymb}	

\newcommand{\e}{\end{equation}}
\newcommand{\bear}{\begin{eqnarray}}
\newcommand{\ear}{\end{eqnarray}}
\newcommand{\hmpc}{{\, h^{-1}\, {\rm Mpc}}}
\def\aj{AJ}
\def\apj{ApJ}
\def\apjs{ApJS}
\def\jcap{JCAP}
\def\mnras{MNRAS}
\def\prl{Physical Review Letters}
\def\aap{A\&A}
\def\prd{Physical Review D}
\def\nat{Nature}
\def\apjs{ApJS}
\def\apjl{ApJ Letters}

\title[Testing isotropy in the 2MRS]{Testing isotropy in the Two
  Micron All-Sky redshift survey with information entropy}
    
\author[Pandey, B.]  {Biswajit Pandey\thanks{E-mail:
    biswap@visva-bharati.ac.in}\\Department of Physics, Visva-Bharati
  University, Santiniketan, Birbhum, 731235, India\\ 
}
       
 \date{\today}

 \pubyear{2016}
  
\begin{document}
\label{firstpage}
\pagerange{\pageref{firstpage}--\pageref{lastpage}}      
\maketitle
       
 \begin{abstract}

We use information entropy to test the isotropy in the nearby galaxy
distribution mapped by the Two Micron All-Sky redshift survey
(2MRS). We find that the galaxy distribution is highly anisotropic on
small scales. The radial anisotropy gradually decreases with
increasing length scales and the observed anisotropy is consistent
with that expected for an isotropic Poisson distribution beyond a
length scale of $90 \, h^{-1}\, {\rm Mpc}$. Using mock catalogues from
N-body simulations, we find that the galaxy distribution in the 2MRS
exhibits a degree of anisotropy compatible with that of the
$\Lambda$CDM model after accounting for the clustering bias of the
2MRS galaxies. We also quantify the polar and azimuthal anisotropies
and identify two directions $(l,b)=(150^{\circ},-15^{\circ})$,
$(l,b)=(310^{\circ},-15^{\circ})$ which are significantly anisotropic
compared to the other directions in the sky. We suggest that their
preferential orientations on the sky may indicate a possible alignment
of the Local Group with two nearby large scale structures. Despite the
differences in the degree of anisotropy on small scales, we find that
the galaxy distributions in both the 2MRS and the $\Lambda$CDM model
are isotropic on a scale of $90 \, h^{-1}\, {\rm Mpc}$.

\end{abstract}

       
       \section{Introduction}

Our current understanding of the Universe relies on a fundamental
assumption that the Universe is statistically homogeneous and
isotropic on sufficiently large scales. It is in general difficult to
prove this assumption in a strictly mathematical sense but it can be
verified using different cosmological observations. Discovery of the
Cosmic Microwave Background Radiation (CMBR) in 1964 \citep{penzias}
and the analysis of the data from the COBE mission launched in 1990
revealed that CMBR has a near uniform temperature across the entire
sky \citep{smoot,fixsen}. This discovery provided possibly the most
powerful evidence for isotropy. Analysis of data from the two
subsequent missions, WMAP launched in 2001 and PLANCK launched in
2009, revealed that CMBR is not completely isotropic. The physics of
CMB anisotropy is now well understood. However various studies
\citep{eriksen, hoftuft,
  akrami,adeplanck1,adeplanck2,schwarz1,land,hanlewis,moss,grupp,dai}
with WMAP and PLACK reported several unexpected features at large
angular scales such as a hemispherical power asymmetry, alignment of
the low multipole moments, a preference for the odd parity modes and
an unexpectedly large cold spot in the southern
hemisphere. Consistency of WMAP and PLANCK results suggest that the
instrumental effects are unlikely to produce such features. So the
present status of the CMBR observations place the assumption of cosmic
isotropy under scanner. The isotropy of the Universe has been favoured
by observations in the other wavelengths such as the X-ray background
\citep{wu,scharf}, the angular distributions of radio sources
\citep{wilson,blake} and Gamma-ray bursts \citep{meegan,briggs}. Some
studies on the distribution of supernovae \citep{gupta,lin}, galaxies
\citep{marinoni, alonso} and neutral hydrogen \citep{hazra} are also
consistent with the assumption of statistical isotropy. However there
are other studies with Type-Ia supernovae
\citep{schwarz2,campanelli,kalus,javanmardi,bengaly}, radio sources
\citep{jackson} and galaxy luminosity function \citep{appleby} which
find the evidence for statistically significant deviation from
isotropy. The signatures of anisotropy can be also detected using the
measurements of large scale bulk flows which are expected to disappear
on large scales. Some studies with peculiar velocity surveys, WMAP
data and x-ray cluster catalog find statistically significant bulk
flows on scales of $100-300 \, h^{-1}\, {\rm Mpc}$ \citep{watkins,
  kashlinsky1, kashlinsky2} whereas some analysis with Type-Ia
Supernovae find no evidence of such bulk flows \citep{huterer}. It is
also important to understand the origin of any observed
anisotropies. There are a number of theoretical studies which predict
the level of statistical anisotropy expected from anisotropic
inflation \citep{barrow, soda} and backreaction of large scale
structure \citep{marrozi}. The Current observational findings suggest
that there is no clear consensus on the issue of cosmic isotropy. It
is important to test the assumption of isotropy in multiple data sets
with different statistical tools. There will be a major paradigm shift
in cosmology if different observations rule out the assumption of
isotropy with high statistical significance.

The present generation of redshift surveys like 2dFGRS \citep{colles},
SDSS \citep{york} and 2MRS \citep{huchra} now provide detailed maps of
the local Universe. The large sky coverage and the large number of
galaxies mapped by these surveys provide an unique opportunity to test
the assumption of isotropy in the local universe using galaxy
distributions. The 2MASS redshift survey (2MRS) \citep{huchra} has
some unique features which make it distinct compared to the previous
optical and far-infrared surveys. The SDSS and 2dFGRS, the two large
redshift surveys have not attempted to be complete over the whole
sky. The 2MASS redshift survey covers the $91\%$ of the entire sky and
selects the galaxies in the near infrared wavelengths around $2 \mu m$
which is less affected by dust extinction and stellar confusion. The
near infrared wavelengths are sensitive to the old stellar populations
which dominate the galaxy masses and thus provides a statistically
uniform galaxy sample in the nearby Universe.  The 2MRS survey is
$97\%$ complete down to the limiting magnitude $K_s=11.75$ and
provides a fair sample of the mass distribution in the local
Universe. These additional features of the 2MRS make it most suitable
for testing isotropy in the local Universe.

\citet{pandey16a} propose an information theory based method for
testing isotropy in a three dimensional distribution. In this paper we
apply this method to the all-sky 2MASS redshift survey to test the
assumption of cosmic isotropy in the nearby Universe.

A brief outline of the paper follows. We describe our method in
Section 2, describe the data in Section 3 and present the results and
conclusions in Section 4.

We have used a $\Lambda$CDM cosmological model with $\Omega_{m0}=0.31$,
$\Omega_{\Lambda0}=0.69$ and $h=1$ throughout.


\section{METHOD OF ANALYSIS}

 The information entropy is originally introduced by Claude Shannon
 \citep{shannon48} in the context of communication of information over
 a noisy channel. In information theory entropy is the key measure
 which quantifies the amount of uncertainty involved in the
 measurement of a random variable. In a more general sense, it gives a
 measure of the amount of information required to describe a random
 variable. For a discrete random variable $X$ with probability
 distribution $P(x)$ and $n$ outcomes $\{X_{i}:i=1,....n\}$, the
 information entropy $H(X)$ associated with $X$ is defined as,
\begin{equation}
H(X) =  - \sum^{n}_{i=1} \, P(X_{i}) \, \log \, P(X_{i})
\label{eq:shannon1}
\end{equation}

\citet{pandey16a} propose a method to test the isotropy of a three
dimensional distribution using information entropy. The method first
requires us to identify a galaxy around which isotropy is to be
tested. The co-ordinates of the rest of the galaxies are then defined
by treating this galaxy as the origin. We carry out an uniform binning
of $\cos\theta$ and $\phi$, where $\theta$ and $\phi$ are respectively
the polar angle and azimuthal angle in spherical polar
co-ordinates. We adopted this binning so as to ensure equal area
$dA=\sin\theta d\theta d\phi$ for each angular bin. If $m_{\theta}$
and $m_{\phi}$ bins are used while binning $\cos\theta$ and $\phi$
respectively, it would result in a total
$m_{total}=m_{\theta}m_{\phi}$ angular bins. An upper limit is imposed
on the radial co-ordinate $r=r_{max}$ as the galaxies are only
available within a finite region. We vary the radial co-ordinate $r$
within this limit and count the number of galaxies within each volume
element defined by the $m_{total}$ angular bins. For a given value of
$r$ each of the element has exactly the same volume
$dv=\frac{r^3}{3}d\Omega$. Let $n_i$ be the number of galaxies
residing inside the $i^{th}$ volume element. A galaxy within a
distance $r$ from the centre can only reside in one of the $m_{total}$
distinct volume elements. One may ask, which particular volume element
a randomly selected galaxy belongs to ? There are a total $m_{total}$
bins and the randomly selected galaxy can reside in any one of
them. The probability of finding the randomly selected galaxy in a
particular bin would depend on the number of galaxies available in
that bin. So we introduce a random variable $X_{\theta\phi}$ with
$m_{total}$ outcomes each given by,
$f_{i}=\frac{n_{i}}{\sum^{m_{total}}_{i=1} \, n_{i}}$ with the
constraint $\sum^{m_{total}}_{i=1} \, f_{i}=1$. Here $f_{i}$ denotes
the probability of finding a randomly selected galaxy in the $i^{th}$
bin. The information entropy associated with $X_{\theta\phi}$ for a
distance $r$ can be written as,
\begin{eqnarray}
H_{\theta\phi}(r)& = &- \sum^{m_{total}}_{i=1} \, f_{i}\, \log\, f_{i} \nonumber\\ &=& 
\log N - \frac {\sum^{m_{total}}_{i=1} \, n_i \, \log n_i}{N}
\label{eq:shannon2}
\end{eqnarray}
Here $N$ is the total number of galaxies within radius $r$. One can
choose any base of the logarithm. We choose the base to be $10$ for
the present work.

In general the probabilities $f_{i}$ will be different for different
elements. The probabilities will have an identical value
$\frac{1}{m_{total}}$ for all the elements only when they are equally
populated. This maximizes the information entropy
$(H_{\theta\phi})_{max}=\log \, m_{total}$ for a given choice of
$m_{\theta}$, $m_{\phi}$ and $r$. We define the relative information
entropy as the ratio of the entropy of a random variable
$X_{\theta\phi}$ to its maximum possible value
$(H_{\theta\phi})_{max}$. The relative Shannon entropy
$\frac{H_{\theta\phi}(r)}{(H_{\theta\phi})_{max}}$ quantifies the
degree of uncertainty in the knowledge of the random variable
$X_{\theta\phi}$. Equivalently we define the residual information
$a_{\theta\phi}(r)=1-\frac{H_{\theta\phi}(r)}{(H_{\theta\phi})_{max}}$
which can be considered as a measure of the degree of anisotropy
present in the distribution. For an isotropic distribution
$H_{\theta\phi}=(H_{\theta\phi})_{max}$ and consequently
$a_{\theta\phi}(r)=0$.

The galaxies are not randomly distributed. The Gravitational
clustering assemble the galaxies into clusters and superclusters which
are linked together in a complex filamentary network surrounded by
empty regions or voids. The present distribution of galaxies are
expected to be highly anisotropic. So the probabilities for different
volume elements will be highly non-uniform. If all the galaxies reside
in a particular volume element then there will be no uncertainty in
identifying the location of the randomly selected galaxy and we will
have $H_{\theta\phi}=0$, $a_{\theta\phi}=1$. This fully determined
situation corresponds to maximum anisotropy and complete lack of
information. On the other hand an uniform probability for all the
$m_{total}$ volume elements make it most difficult and uncertain to
predict the location of a randomly selected galaxy. This maximizes the
information entropy to $H_{\theta\phi}=\log \, m_{total}$ turning
$a_{\theta\phi}=0$. This corresponds to a distribution which is
completely isotropic with maximum amount of information. The galaxy
distribution is expected to be anisotropic on small scales but
eventually should reach isotropy with increasing size of the angular
bins $dA$ and radius $r$ given the assumption of isotropy holds on
large scales. We change the value of $r$ starting from a small radius
and gradually increase it in uniform steps upto its maximum value
$r_{max}$. We study how $a_{\theta\phi}(r)$ varies with $r$ for a
given choice of $m_{\theta}$ and $m_{\phi}$.

Following the definition of the radial anisotropy $a_{\theta\phi}(r)$
one can also define similar measures for the polar anisotropy
$a_{\phi}(\theta)=1-\frac{H_{\phi}}{(H_{\phi})_{max}}$ and the
azimuthal anisotropy
$a_{\theta}(\phi)=1-\frac{H_{\theta}}{(H_{\theta})_{max}}$ as function
of $\theta$ and $\phi$. For this one needs to carry out the sum over
$m_{\phi}$ or $m_{\theta}$ instead of $m_{total}$ in
\autoref{eq:shannon2}. It should be noted that in this case $N$ in
\autoref{eq:shannon2} would be the total number of galaxies inside all
the $m_{\phi}$ or $m_{\theta}$ volume elements available at different
$\theta$ or $\phi$ respectively. The polar anisotropy
$a_{\phi}(\theta)$ quantify the anisotropy among all the $\phi$ bins
at each $\theta$ value. Similarly the azimuthal anisotropy
$a_{\theta}(\phi)$ quantify the anisotropy among all the $\theta$ bins
at each $\phi$ value. We fix the value of $r$ at $r_{max}$ or any
desired radius and determine $a_{\phi}(\theta)$ and $a_{\theta}(\phi)$
at different $\theta$ and $\phi$ values respectively. Note that one
can also define similar anisotropy measures $a_{\phi}(r)$ and
$a_{\theta}(r)$ and measure them as a function of radius $r$ at fixed
$\theta$ and $\phi$ respectively. However in the present work we only
consider $a_{\theta\phi}(r)$, $a_{\phi}(\theta)$ and
$a_{\theta}(\phi)$ to characterize the anisotropies present in the
galaxy distribution. We use the galactic co-ordinates $(l,b)$
throughout the present analysis. So we replace $\theta$ and $\phi$ in
the previous definitions by $b$ and $l$ respectively.

The present day galaxy distribution is highly non-Gaussian. The
information entropy is related to the higher order moments of a
distribution \citep{pandey16b}. So it captures more information about
the probability distribution and may serve as a better measure for
anisotropies present in a distribution. We would also like to point
out here that these measures of anisotropy would never be exactly zero
and would be also sensitive to binning and sub-sampling. This relative
character of information does not pose any difficulty as it only
signifies that the magnitudes of the information is defined with
respect to a reference frame and comparison in anisotropies should be
always done in the same reference frame. Here we adopt a working
definition of isotropy. We compare the anisotropies measured in a
distribution with that from Poisson distributions and consider the
distribution to be isotropic only when the measured anisotropy lies
within the $1-\sigma$ errorbars of the anisotropy expected from a
Poisson distribution. We use the same binning and sampling to compute
the anisotropies in both the distributions. However for large number
of pixels, the size of the volume elements are smaller and
consequently the shot noise in an anisotropic distribution may persist
upto a larger length scale as compared to a homogeneous and isotropic
Poisson random distribution. Our previous definition of isotropy may
not apply in such situations. In such cases we consider a distribution
to be isotropic when the rate of change of the degree of radial anisotropy
$a_{\theta\phi}(r)$ with $r$ decreases to nearly zero.

\begin{figure*}
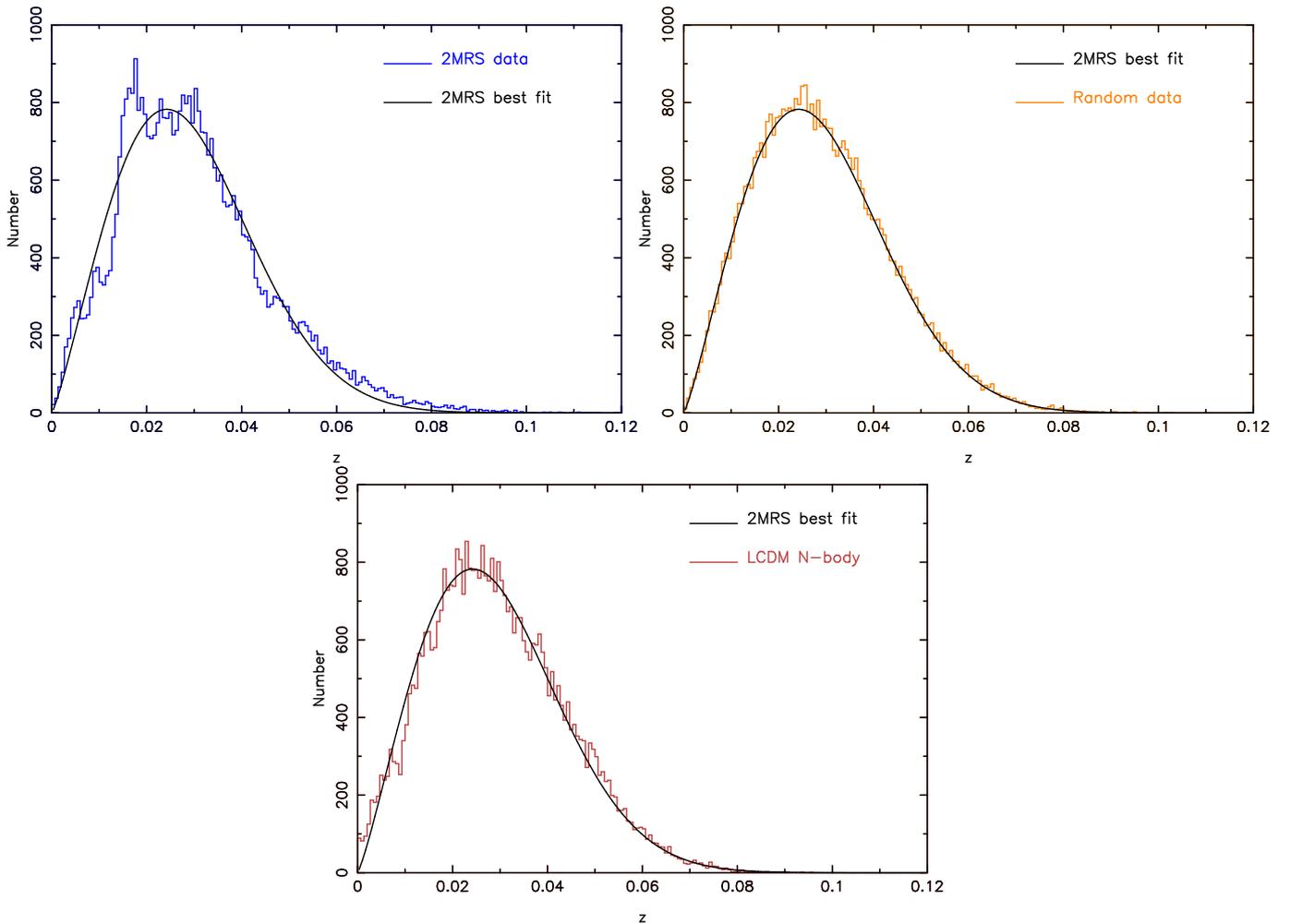

\resizebox{9cm}{!}{\rotatebox{-90}{\includegraphics{plot1.ps}}}%
\resizebox{9cm}{!}{\rotatebox{-90}{\includegraphics{plot2.ps}}}\\
\resizebox{9cm}{!}{\rotatebox{-90}{\includegraphics{nbody1.ps}}}%
\caption{The top left panel shows the redshift histogram in the 2MRS
  catalog in the redshift range $0 \leq z \leq 0.12 $ using uniform
  binsize $cz=200 km/s$. A least square fit (\autoref{eq:fit}) to the
  data is also plotted together. The best fit redshift distribution is
  then used to simulate the mock catalogues from Poisson distributions
  and N-body simulations. We show the redshift histograms in a mock
  catalogue from a Poisson random distribution and N-body simulation
  of the $\Lambda$CDM model in the top right panel and the bottom
  panel respectively.}
  \label{fig:selfunc}
\end{figure*}

\section{DATA}

\subsection{THE 2MRS CATALOGUE}

The Two Micron All Sky Redshift Survey (2MRS) \citep{huchra} provides
a three dimensional distribution of $\sim 45,000$ galaxies in the
nearby Universe. The final survey covers $91\%$ of the sky and is
$97.6\%$ complete to a limiting magnitude of $K_{s}=11.75$. We
download the 2MRS catalogue from http://tdc-www.harvard.edu/2mrs/.
The catalog contain $43,533$ galaxies with apparent infrared magnitude
$K_{s} \leq 11.75$ and colour excess $E(B-V) \leq 1$ in the region
$|b| \geq 5^{\circ}$ for $30^{\circ} \leq l \leq 330^{\circ}$ and $|b|
\geq 8^{\circ}$ otherwise. We calculate the absolute magnitudes of
these galaxies by using their extinction corrected apparent magnitude
in the $K_{s}$ band and applying the k-correction $k(z)=-6\log(1+z)$
\citep{kochanek} and e-correction $e(z)=3.04z$ \citep{branchini} to
account for the k-correction and evolutionary correction in the
luminosity of the galaxies.

The 2MRS catalog is flux limited. We extract volume limited samples
from the 2MRS catalog but find that they are quite sparse and are
unsuitable for the present analysis. So we do not apply any absolute
magnitude cut but restrict our sample to $z\leq 0.12$ beyond which
there are very few galaxies. The redshift limit is required to
simulate the mock catalogues from Poisson random distributions and
N-body simulations. We finally have $43,305$ galaxies in our 2MRS
sample. We use this flux limited sample for the present analysis.

\subsection{REDSHIFT DISTRIBUTION IN THE 2MRS}

We calculate the redshift distribution of the 2MRS galaxies out to
$cz=36000 \, km/s$ using uniform binsize of $200 \, km/s$. The
redshift distribution for the 2MRS sample is shown in the left panel
of \autoref{fig:selfunc}. We model the redshift distribution using a
parametrized fit \citep{erdogdu2, erdogdu1} given by,
\begin{equation}
\frac{dN(z)}{dz}= A \,z^{\gamma} \, \exp[-\big{(\frac{z}{z_{c}})}^\alpha]
\label{eq:fit}
\end{equation}
We find the best fit parameters $A = 116000 \pm 5100$, $\gamma = 1.188
\pm 0.093$, $z_{c} = 0.031 \pm 0.002$ and $\alpha = 2.059 \pm 0.149$
by fitting the above equation to the 2MRS redshift distribution. The
best fit is shown with a smooth solid line in the top left panel of
\autoref{fig:selfunc}. When normalized, the \autoref{eq:fit} gives the
probability $P(z)$ of detecting a galaxy at redshift $z$.

\subsection{RANDOM CATALOGUES}

We simulate a set of mock random catalogues for the 2MRS galaxies
using Monte Carlo simulation. The maxima of the function in
\autoref{eq:fit} is at $z=z_{c}
(\frac{\gamma}{\alpha})^\frac{1}{\alpha}$. Plugging the best fit
values of the parameters $A$, $\gamma$, $z_{c}$ and $\alpha$ yields
the maxima at $z_{max}=0.0242$. We calculate the maximum probability
$P_{max}$ by plugging the value of $z_{max}$ in \autoref{eq:fit}. We
randomly choose a redshift in the range $0 \leq z \leq 0.12$ and a
probability value is randomly chosen in the range $0 \leq P(z) \leq
P_{max}$. We calculate the actual probability of detecting the galaxy
at the selected redshift using \autoref{eq:fit} and compare it to the
randomly selected probability value. If the calculated probability
value is larger than the randomly selected probability value, the
randomly selected redshift is accepted and assigned isotropically
selected galactic co-ordinates $l$ and $b$ within the sky coverage of
the 2MRS survey. We simulate $30$ such random mock catalogues. Each
random catalogue contains exactly the same number of galaxies
distributed over the same region as the actual 2MRS survey. By
construction, the random mock catalogues are isotropic around us and
they have exactly the same selection function as the 2MASS redshift
survey. For comparison, the redshift distribution in a mock random
catalogue is shown together with the best fit to the 2MRS data in the
top right panel of \autoref{fig:selfunc}.

\subsection{MOCK CATALOGUES FROM N-BODY SIMULATIONS}

We simulate the distributions of dark matter in the $\Lambda$CDM model
using a Particle-Mesh (PM) N-Body code. We use $256^{3}$ particles on
a $512^{3}$ mesh to simulate the present day distribution of dark
matter in a comoving volume of $(921.6 \, h^{-1}\, {\rm Mpc})^3$. The
following values of the cosmological parameters are used in the
simulations: $\Omega_{m0}=0.31$, $\Omega_{\Lambda0}=0.69$, $h=0.68$,
$\sigma_{8}=0.81$ and $n_{s}=0.96$ \citep{adeplanck3}. The simulations
were run for three different realizations of the initial density
fluctuations. We treat the individual particles as galaxies and place
an observer at the centre of these boxes. We use the peculiar
velocities to produce the galaxy distributions in redshift space and
extract $10$ mock catalogues for the 2MRS galaxy distribution from
each boxes following the same method as used for random
distribution. The only difference is that here we do not need to
generate isotropically selected galactic co-ordinates for the mock
galaxies as these are decided by their co-ordinates. We finally have
$30$ mock catalogues for the 2MRS galaxies from the N-body simulations
of the $\Lambda$CDM model.

\begin{figure*}
\resizebox{9cm}{!}{\rotatebox{-90}{\includegraphics{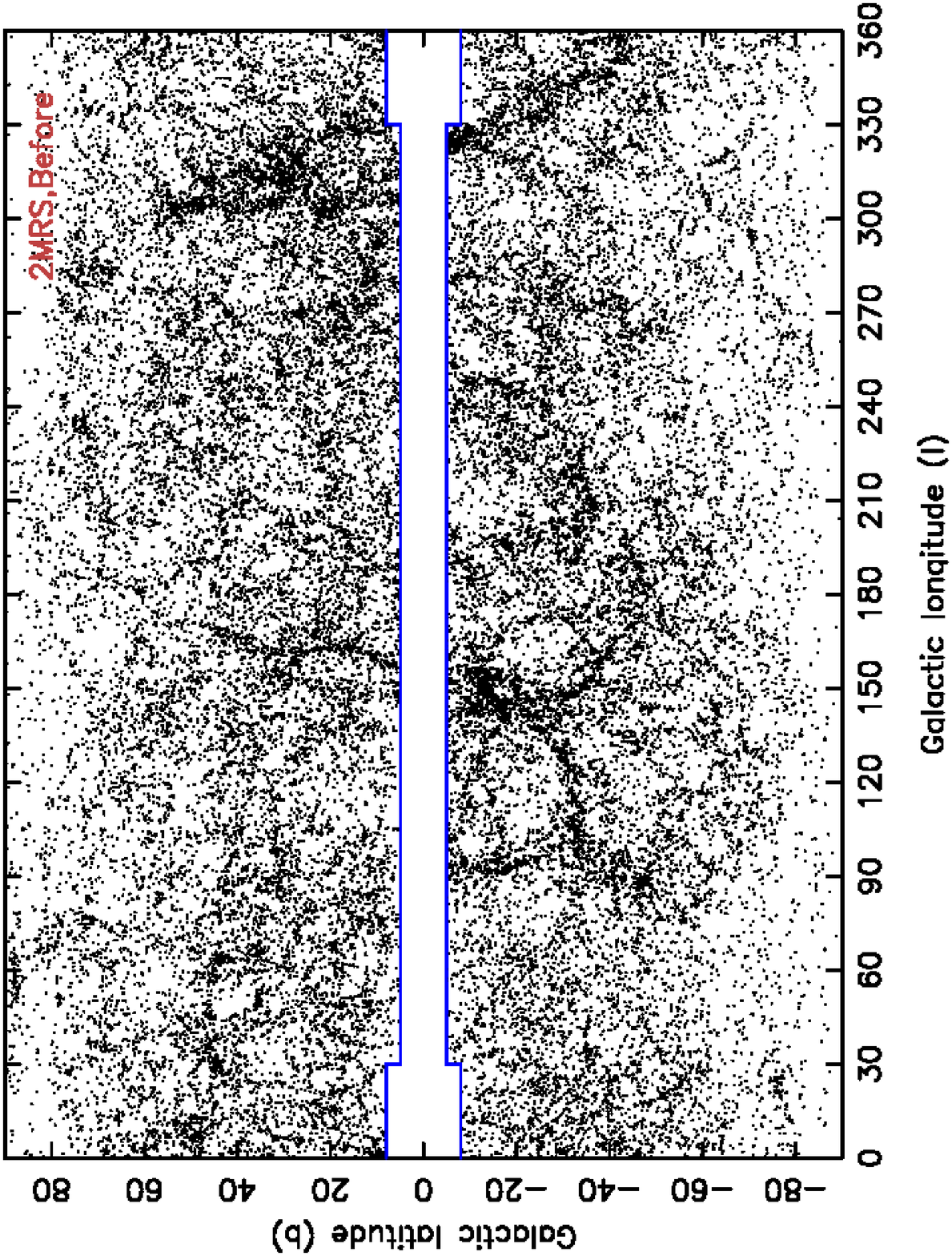}}}%
\resizebox{9cm}{!}{\rotatebox{-90}{\includegraphics{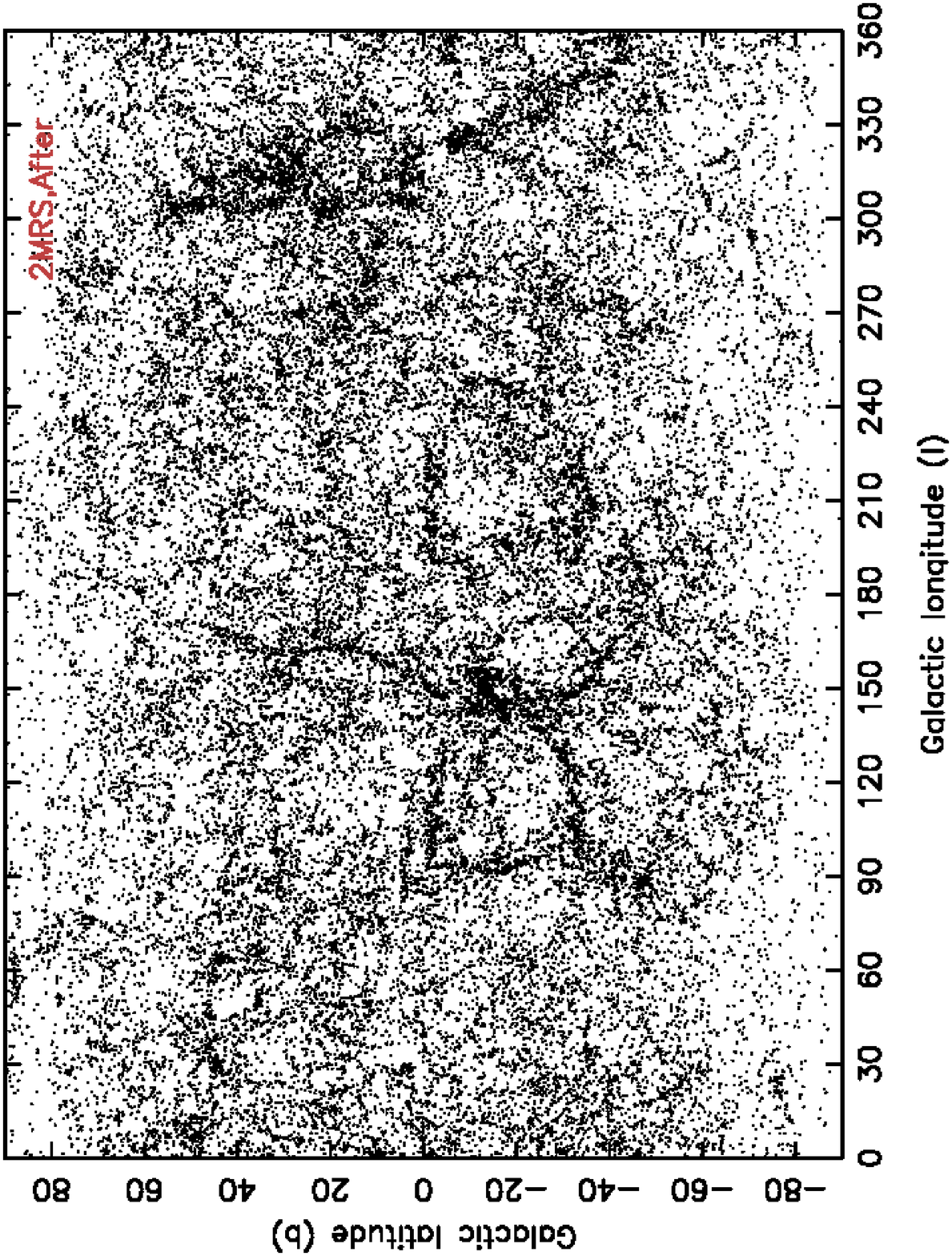}}}\\
\resizebox{9cm}{!}{\rotatebox{-90}{\includegraphics{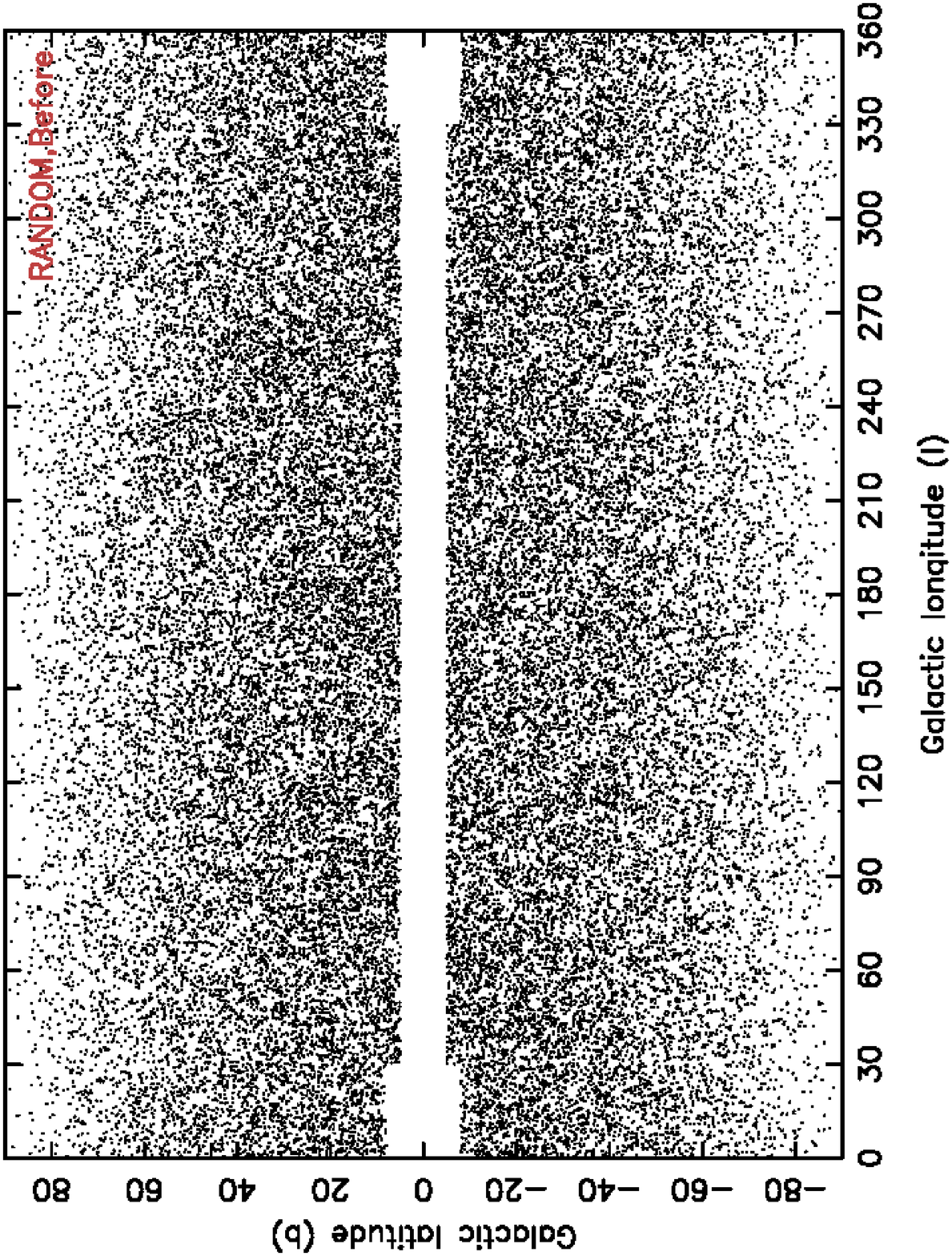}}}%
\resizebox{9cm}{!}{\rotatebox{-90}{\includegraphics{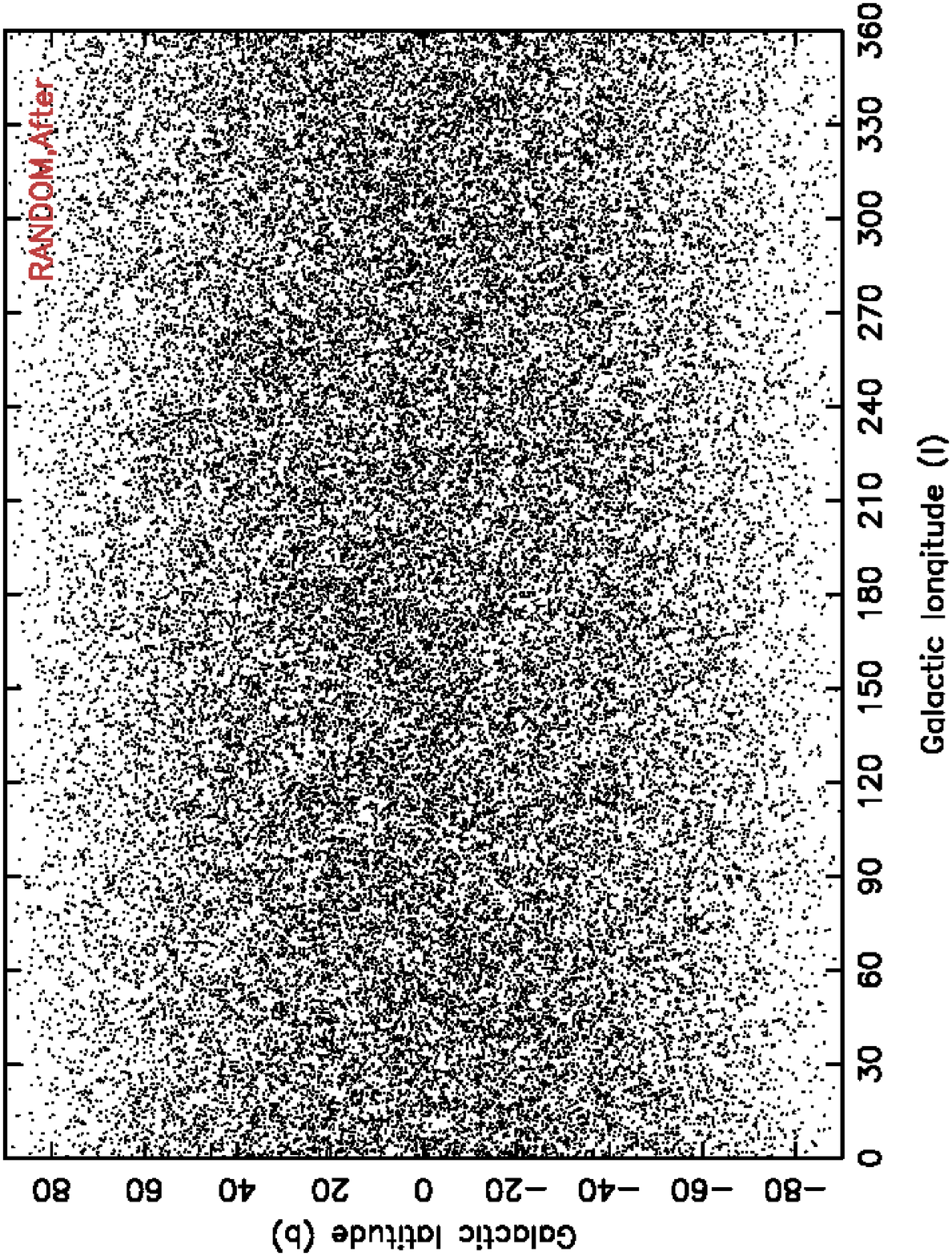}}}\\
\resizebox{9cm}{!}{\rotatebox{-90}{\includegraphics{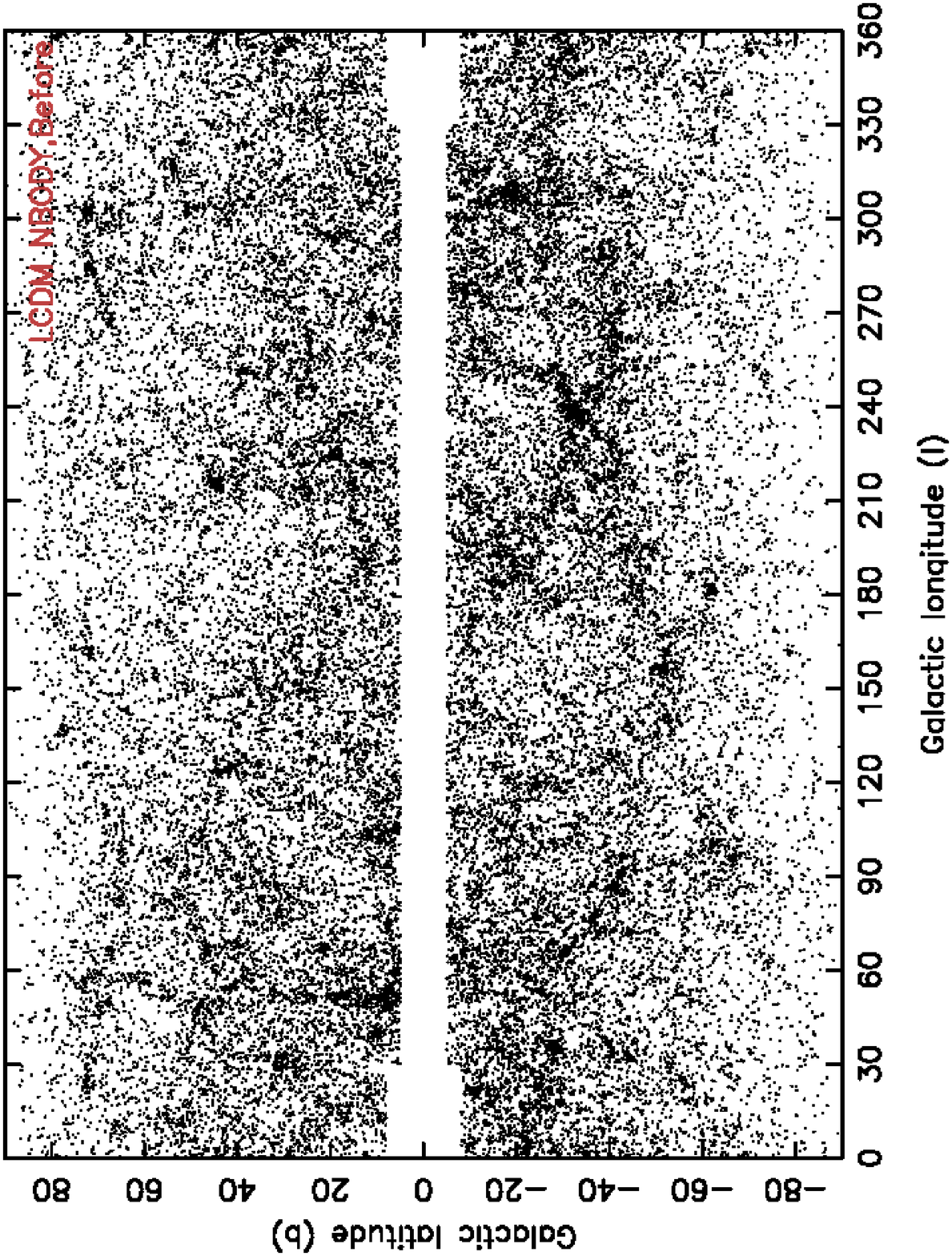}}}%
\resizebox{9cm}{!}{\rotatebox{-90}{\includegraphics{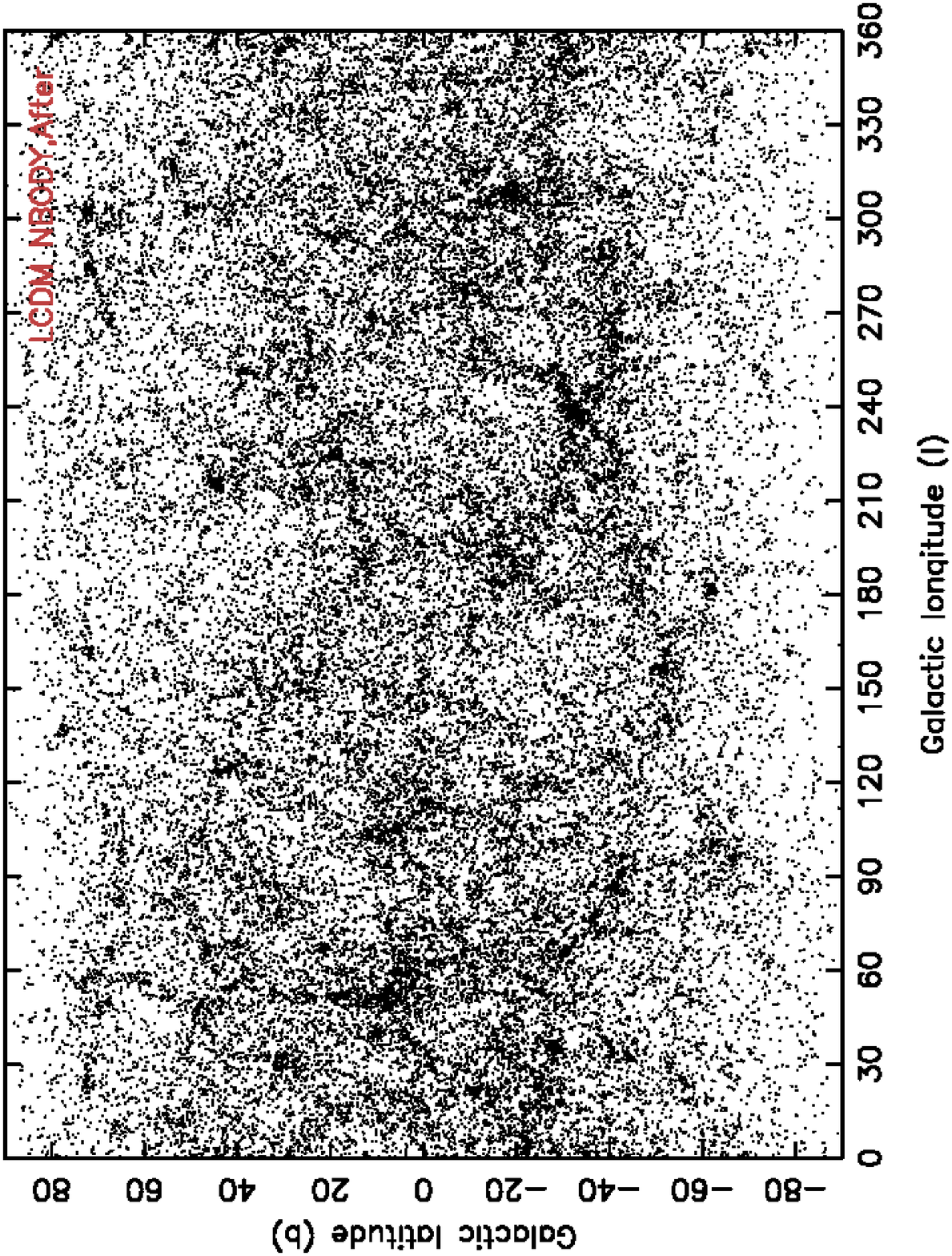}}}\\
\caption{The top left, middle left and bottom left panels show the
  galactic co-ordinates of the galaxies before filling the ZOA in the
  2MRS, a random mock catalogue and a mock catalogue from the
  $\Lambda$CDM N-Body simulation respectively. The top right, middle
  right and bottom right panels show the same after filling the ZOA.
}
  \label{fig:zoa}
\end{figure*}

\subsection{FILLING THE ZONE OF AVOIDANCE}

The Zone of Avoidance (ZOA) is the region of the sky near the Galactic
plane where the observations are obscured due to the extinction by
Galactic dust. The selection in the near infra-red in the 2MRS reduces
the impact of the zone of avoidance. \citet{huchra} selected 45,086
2MRS sources which has apparent infrared magnitude $K_{s} \leq 11.75$
and colour excess $E(B-V) \leq 1$ in the region $|b| \geq 5^{\circ}$
for $30^{\circ} \leq l \leq 330^{\circ}$ and $|b| \geq 8^{\circ}$
otherwise. They further rejected the sources which are of galactic
origin (multiple stars, planetary nebulae, HII regions) and also
discarded the sources which are in regions of high stellar density and
absorption. After correction for these systematic effects, the final
2MRS catalog provided by \citet{huchra} contain 43,533 galaxies. We
use the final 2MRS catalog and the distribution of the 2MRS galaxies
in galactic co-ordinates is shown in the top left panel of
\autoref{fig:zoa}.  For a comparison, the distribution for a mock
catalogue from random distribution and N-body simulation are also
shown in the middle and bottom left panels of the same figure. The ZOA
can be clearly identified at the middle of these distributions. The
present study aims to explore the isotropy in the observed mass
distribution in the local Universe and it would be desirable to have
the galaxy distribution over the full sky. We fill the ZOA by randomly
cloning individual galaxies from above and below the ZOA (the unmasked
region) and then shifting them in latitude to random locations in the
masked region so that it finally has the same average density of
galaxies as the unmasked region \citep{lyndenbell}. Although this
fails to interpolate the large scale structures across the ZOA, it
serves the purpose of constructing a full sky three dimensional galaxy
distribution without introducing any spurious signals of
anisotropy. We finally have $4,375$ clones filling the ZOA. Our 2MRS
sample contains $43,305$ galaxies before filling the ZOA and contains
a total $47,680$ galaxies after filling it. Following the same
procedure we fill up the ZOA in all the mock catalogues from Poisson
random distributions and N-body simulations. After filling the ZOA,
the distribution of galaxies in the 2MRS catalogue, a mock random
catalogue and a mock catalogue from N-body simulation are shown in the
top right, middle right and bottom right panel of \autoref{fig:zoa}
respectively.

We note that in reality the zone of avoidance is not as symmetric as
defined in \citet{huchra} but as we keep the cloned galaxies in the
masked regions and carry out our analysis in coordinate space, we do
not expect these to influence our results. We have also checked that
filling the zone of avoidance by uniform strips or chunks instead of
individual clones does not make any difference to our results. Finally
it may be noted that filling the ZOA is not necessary if one limits
the analysis to the unmasked regions outside the ZOA.

\subsection{JACKKNIFE SAMPLES FOR THE 2MRS}

We have only one galaxy sample from the 2MRS. We prepare $30$
jackknife samples from the 2MRS data to estimate the errorbars for our
measurements. The final 2MRS sample used in this analysis contain
$47,680$ galaxies after the ZOA is filled with clones. We used the
general delete-m observations jackknife method for equal m. For each
of the $30$ jackknife samples, we randomly omit $12,680$ galaxies
from the final 2MRS sample. This provides us $30$ jackknife samples
for the 2MRS each containing $35,000$ galaxies.

\begin{figure*}
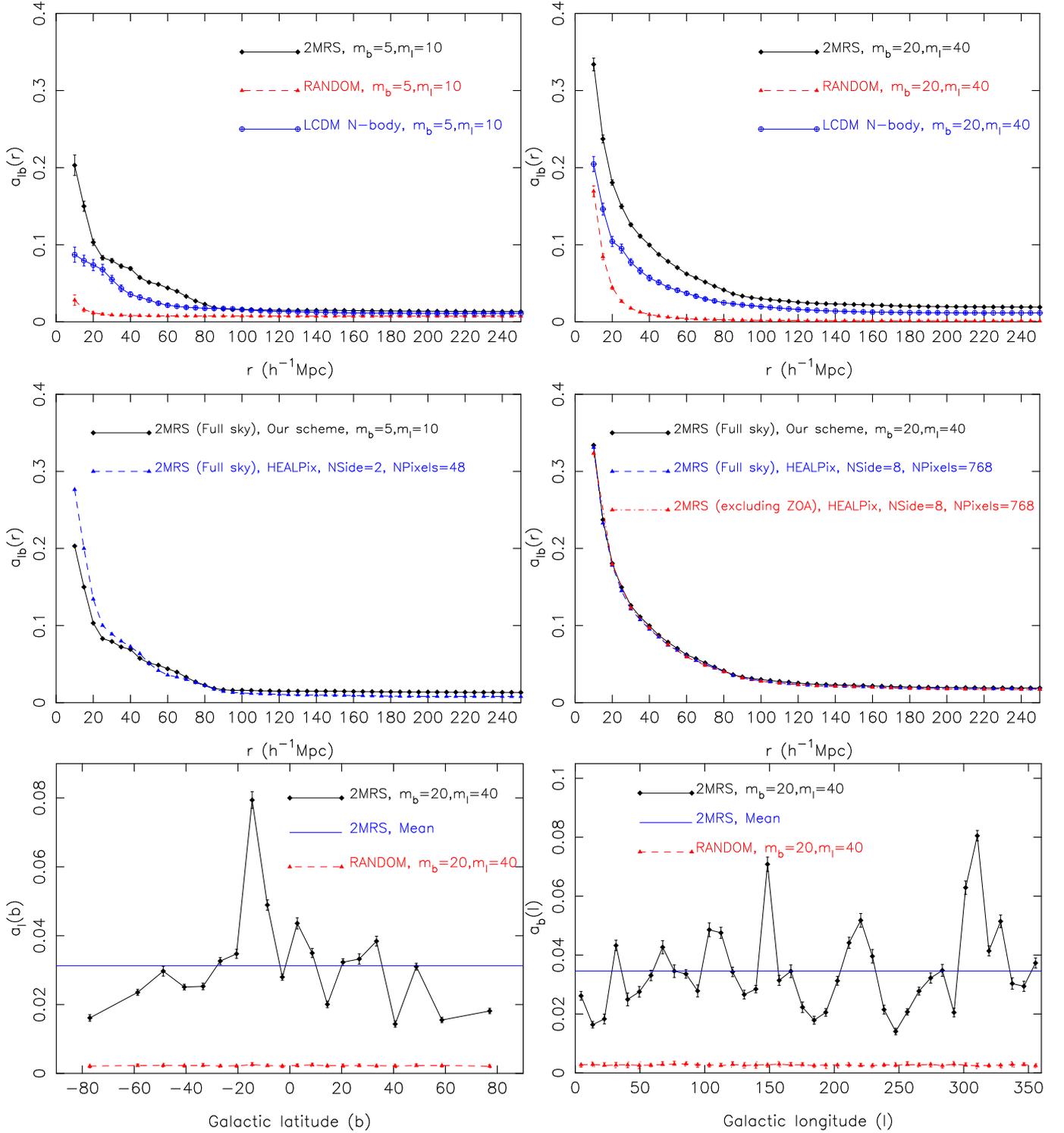

\resizebox{9cm}{!}{\rotatebox{-90}{\includegraphics{plot7.ps}}}%
\resizebox{9cm}{!}{\rotatebox{-90}{\includegraphics{plot9.ps}}}\\
\resizebox{9cm}{!}{\rotatebox{-90}{\includegraphics{hpix1.ps}}}%
\resizebox{9cm}{!}{\rotatebox{-90}{\includegraphics{hpix2.ps}}}\\
\resizebox{9cm}{!}{\rotatebox{-90}{\includegraphics{plot13.ps}}}%
\resizebox{9cm}{!}{\rotatebox{-90}{\includegraphics{plot15.ps}}}\\
\caption{ The top left panel shows the radial anisotropy $a_{lb}(r)$
  in the 2MRS, Poisson distributions and N-body simulations as a
  function of length scales for bin sizes of $m_{b}=5$ and
  $m_{l}=10$. The top right panel shows the same but for a different
  choice of the number of bins, $m_{b}=20$ and $m_{l}=40$. The middle
  left and right panels compares the radial anisotropy $a_{lb}(r)$
  measured using our scheme and the HEALPix software for
  $(m_{b}=5,m_{l}=10)$ and $(m_{b}=20,m_{l}=40)$ respectively. The
  bottom left and the bottom right panels respectively show the polar
  anisotropy $a_{l}(b)$ and the azimuthal anisotropy $a_{b}(l)$ for
  the 2MRS galaxies and the mock Poisson random samples. The mean
  levels of polar and azimuthal anisotropies in the 2MRS sample are
  also shown in the two bottom panels. The $1\sigma$ errorbars shown
  for the 2MRS galaxies in different panels are obtained from $30$
  jackknife samples and the $1\sigma$ errorbars for both the Poisson
  samples and N-body simulations are obtained from $30$ respective
  mock catalogues.}

  \label{fig:isolbr}
\end{figure*}

\section{RESULTS AND CONCLUSIONS}

We show the degree of radial anisotropy $a_{lb}(r)$ as a function of
length scales $r$ for the 2MRS sample and the mock samples from
Poisson random distributions and N-body simulations of the
$\Lambda$CDM model in the top two panels of \autoref{fig:isolbr}. The
choice of the number of bins are indicated in each panel. The
$1\sigma$ error bars for the 2MRS galaxies are obtained from $30$
jackknife samples. The $1\sigma$ error bars for the random data and
N-body data are obtained from the $30$ respective mock samples. The
top left panel shows the results for $m_{b}=5$ and $m_{l}=10$. This
divides the entire sky into $50$ patches with exactly identical
area. We see that the radial anisotropy in the 2MRS is maximum at the
smallest radius considered. At the radius of $10 \hmpc$ the galaxy
distribution is highly anisotropic and the level of anisotropy is much
larger than that observed in the mock random samples. The mock random
samples are isotropic by construction and the small anisotropies
observed in the random samples are purely an outcome of the shot
noise. It is interesting to note that the degree of the radial
anisotropy in the 2MRS galaxy distribution decreases with increasing
length scales and it reaches a plateau at $\sim 90 \hmpc$ where it is
consistent with the level of anisotropies expected in a Poisson random
distribution.  In the top right panel we show the same results but for
a different choice of the number of bins. Here we choose $m_{b}=20$
and $m_{l}=40$ i.e. the entire sky is now divided into $800$ equal
area patches resulting into that many radial volume
elements. Expectedly the level of anisotropies shoot up at small scale
in both the 2MRS and random samples due to increase in the shot
noise. The contribution of shot noise to the observed anisotropy is
expected to dominate on small scales and it becomes negligible on
large scales. We see the same trend for the 2MRS galaxies in the top
right panel where the observed anisotropy decreases with increasing
length scales and reaches a plateau at $\sim 90 \hmpc$. However the
differences in the degree of anisotropies in the 2MRS and the random
samples increase when larger number of bins are used.

The present day galaxy distribution is highly anisotropic on small
scales and the observed anisotropies originate from gravitational
clustering, redshift space distortions \citep{kaiser}, shot noise and
the selection effects. We maintain the same shot noise and selection
function in the random samples and the 2MRS sample. So the differences
in the anisotropies in the galaxy distribution and the random samples
are primarily due to the clustering of galaxies and the redshift space
distortions.

The galaxies are distributed in the filamentary cosmic web. The galaxy
clusters are usually located at the nodes where the filaments
intersect. The virialized bound structures such as galaxy clusters are
elongated along the line of sight in redshift space due to the random
velocity dispersions which is popularly known as Fingers of God (FOG)
effect. The individual volume elements are larger when a smaller
number of pixels are used. Consequently they are expected to host a
statistically similar number of these structures given the isotropy of
the matter distribution. The increase in the number of bins
corresponds to a decrease in the transverse dimension of the volume
elements leading to a decrease in their volumes for a given radial
extension. This makes it less likely to have a statistically similar
number of FOGs in the different volume elements. But eventually this
would happen at a large radius where the volume of the individual
elements becomes significantly larger. So these differences in the
anisotropies between the galaxy distribution and the Poisson
distribution are the outcome of non-linear gravitational clustering on
that angular scales.  Despite these differences, the similarity in the
results in the top two panels of the \autoref{fig:isolbr} suggest that
the galaxy distribution in the 2MRS become isotropic on length scales
beyond $\sim 90 \hmpc$.

We also compare the radial anisotropy expected in the $\Lambda$CDM
model to that observed in the 2MRS in the top two panels of
\autoref{fig:isolbr}. We see that the mock catalogues from the
$\Lambda$CDM N-Body simulations exhibit a lower degree of anisotropy
as compared to the 2MRS galaxies. This indicates that the 2MRS
galaxies in the $K_{s}$ band are a biased tracer of the underlying
mass distribution. A biased distribution is expected to be more
anisotropic as compared to an unbiased distribution due to their
differences in clustering. Using the magnitude of the clustering
dipole \citet{maller} find that the 2MASS galaxies in the $K_{s}$ band
have a linear bias of $b\sim1.4$. A further analysis of cosmological
large scale flows \citep{davis} also suggest that the 2MRS galaxies
have a linear bias of $b\sim1.4-1.5$. So the differences in the degree
of anisotropy in our analysis is most likely related to the fact that
the 2MRS galaxies in the $K_{s}$ band are not an unbiased tracer of
the mass distribution. However it is interesting to note that despite
the differences in the degree of anisotropy between the galaxy
distribution in 2MRS and $\Lambda$CDM model, both the distributions
become isotropic beyond a length scales of $90 \hmpc$. We shall
address the dependence of anisotropy on the clustering bias and
explore the possibility of determining the linear bias from it in a
separate work (in preparation).

The present analysis uses a tiling strategy which provides equal area
pixelization of the sky but the pixel shapes may be quite different
around the poles and the equator. This may have an important effect in
the measured anisotropy. To examine this further, we use the HEALPix
software \citep{gorski1,gorski2} to calculate the radial anisotropy in
the 2MRS data as function of $r$ and compare them with the
measurements from our method. The left and right middle panels of
\autoref{fig:isolbr} compares the measurements from HEALPix and our
scheme. In HEALPix we have used (NSide$=2,$ NPixels$=48$) and
(NSide$=8,$ NPixels$=768$) to compare the results with that from our
method for $(m_{b}=5,m_{l}=10)$ and $(m_{b}=20,m_{l}=40)$
respectively. We find that the measured anisotropies by HEALPix and
our scheme show some differences when a small number of pixels are
used. But interestingly they provide nearly identical results when the
number of pixels are increased. This is most possibly related to the
fact that the transverse dimensions of the volume elements become
negligible as compared to their radial dimensions when a larger number
of pixels are used. The statistics employed in this work use the
number counts in the different volume elements integrated along the
radial direction. So the disparity between the shape of the pixels
becomes less important as compared their volumes when the number of
pixels are increased. On the other hand, the differences between the
radial and transverse dimensions of the volume elements are smaller
when a smaller number of pixels are used and consequently the
variations in their shapes may affect the anisotropy measurements. We
also compute the radial anisotropy using only the galaxies outside the
ZOA and compare it with the results obtained from the full sky
analysis. We use HEALPix to analyze the unmasked regions of the sky
using NSide$=8,$ NPixels$=768$ and find that the measured anisotropies
are exactly identical in both the cases. The comparison is shown in
the middle right panel of \autoref{fig:isolbr}. This indicates that
the anisotropy measure employed here is quite robust against
incomplete sky coverage and reliably captures the anisotropic
character of the distribution.

We show the polar anisotropy $a_{l}(b)$ as a function of the galactic
latitudes $b$ for the 2MRS galaxy sample and the mock random samples
in the bottom left panel of \autoref{fig:isolbr}. We use $m_{b}=20$
and $m_{l}=40$ for this analysis. Both the 2MRS and the random samples
exhibit a small degree of polar anisotropy. However the degree of
polar anisotropy in the 2MRS galaxy distribution is noticeably higher
as compared to the random samples. The anisotropy curve for the 2MRS
galaxies is also spiky and irregular due to the anisotropies present
in the galaxy distribution. The mean polar anisotropy for the 2MRS
galaxies is also shown in the same plot. We see a distinct spike in
the polar anisotropy curve for the 2MRS galaxies at $b=-15^{\circ}$
where the polar anisotropy is $\sim 250\%$ of its mean value.

The azimuthal anisotropy $a_{b}(l)$ as a function of galactic
longitude $l$ for the 2MRS galaxy sample and the mock random samples
are shown in the bottom right panel of \autoref{fig:isolbr}. The
number of bins used in the analysis are $m_{b}=20$ and $m_{l}=40$. We
find a small degree of azimuthal anisotropy in both the 2MRS and the
random samples. The degree of azimuthal anisotropy in the random
samples are smaller than the 2MRS galaxy sample. The anisotropy curve
for the random sample appears much smoother than the 2MRS galaxy
sample. The random samples are isotropic by construction and the smaller
amount of anisotropies observed in them arise purely due to the
discreteness noise. On the other hand, the galaxy distribution is
anisotropic due to the presence of large scale structures. This
accounts for the relatively higher degree of anisotropy and the
irregular nature of the anisotropy curve in the 2MRS sample compared
to the mock random samples. The mean azimuthal anisotropy in the 2MRS
sample is also shown together in the same plot. We find two distinct
spikes in the azimuthal anisotropy curve, one at $l=150^{\circ}$ and
another at $l=310^{\circ}$ where the anisotropies are $205\%$ and
$235\%$ of the mean azimuthal anisotropies in the 2MRS.

We repeat the analysis with somewhat larger number of bins and find
that the spikes in the two bottom panels of \autoref{fig:isolbr}
appear roughly at the same locations irrespective of the choice of the
number of bins. But the results become shot noise dominated when a
very large number of bins are used. The observed spikes in the
anisotropy curves are most likely produced by some prominent large
scale structures in the nearby Universe. One can see two visibly
distinct structures at $(l,b)=(150^{\circ},-15^{\circ})$ and
$(l,b)=(310^{\circ},-15^{\circ})$ in the top two panels of
\autoref{fig:zoa}. It is interesting to note that the distinct spikes
in the anisotropies in the bottom two panels of \autoref{fig:isolbr}
appear at the same locations. Although it is difficult to reliably
distinguish the large scale structures in projection, it is still
interesting to note further that the two other smaller spikes which
appears at $l=120^{\circ}$ and $l=220^{\circ}$ in the bottom right
panel of \autoref{fig:isolbr} correspond to two apparently underdense
regions at those locations in the top right panel of
\autoref{fig:zoa}.

The Doppler effect due to the relative motion between the earth and
the CMB rest frame is known to introduce a dipole anisotropy in the
temperature of the CMB. The motion of the Local Group containing our
galaxy is caused by the large scale structures in its
neighbourhood. Analysis of the COBE DMR and PLANCK data suggest that
the Local Group is moving with a velocity $\sim 600 km/s$ toward
$(l,b)=(276^{\circ},30^{\circ})$ whereas the CMB dipole anisotropy is
observed towards $(l,b)=(264^{\circ},48^{\circ})$ \citep{kogut,
  aghanim}. The misalignment of the CMB dipole with the clustering
dipole and its convergence has been addressed in many studies
\citep{yahil, lyndenbell, rowan, erdogdu3, lavaux, bilicki}. There is
still no consensus on this issue mainly due to the sparseness of data
at very large distances.

Clearly, the spikes observed in the two bottom panels of
\autoref{fig:isolbr} are not in the same direction as the CMB dipole
or the clustering dipole. It is interesting to note that the two
spikes in the azimuthal anisotropy $a_{b}(l)$ are separated by
$160^{\circ}$ i.e. they lie roughly in two opposite sides of the
$b=-15^{\circ}$ cone. The small value of $b$ combined with the fact
that the two most anisotropic directions lie opposite each other in
$l$ indicates a possible alignment of the Local Group with two nearby
large scale structures. The radial anisotropy $a_{lb}(r)$ cease to
exist beyond $r=90 \hmpc$ which imply that these large scale
structures must lie within this radius. We verify this by changing
$r_{max}$ from $250 \hmpc$ to $90 \hmpc$ and repeating the
analysis. We again find the two prominent spikes in the anisotropy
curves at the same locations as noticed earlier. It may be noted that
some earlier studies with observed data \citep{peebles, zitrin} and
simulations \citep{klypin} reported that the Local Group may reside in
a filament. It has been also suggested by some studies \citep {tully,
  mccall} that the Local Group is a part of the Local Sheet which
surrounds the Local Void \citep{tully1}. It is interesting to note
that using 2MASS photometric redshift measurements, \citet{kovac}
reported a significant antipodal anisotropy in the direction of the
CMB cold spot $(l,b=209^{\circ},-57^{\circ})$. They find that the
Eridanus supervoid reaches our closest vicinity in the directions of
the CMB Cold Spot and continues to the nearby antipodal directions
$(l,b=29^{\circ},57^{\circ})$ traversing upto the Northern Local
Supervoid beyond which the antipodal line of sight becomes overdense
due to the presence of Hercules and Corona Borealis superclusters.

 Our analysis indicates that the nearby universe is highly
anisotropic. The observed anisotropy gradually decreases with
increasing radial distance and the galaxy distribution in the 2MASS
redshift survey becomes statistically isotropic beyond a length scales
of $90 \hmpc$. In this study we have used the flux limited 2MRS sample
and tested the isotropy of the Universe only around us. In future we
plan to use the 2MASS photometric redshift catalogue \citep{bilicki1}
to construct volume limited samples for our study which will enable us
to address the isotropy around different galaxies in the
Universe. While searching for anisotropy in the polar and azimuthal
directions, we identify two directions
$(l,b)=(150^{\circ},-15^{\circ})$ and
$(l,b)=(310^{\circ},-15^{\circ})$ which are significantly anisotropic
compared to the other directions in the sky. Their preferential
orientations may indicate a possible alignment of the Local Group with
two nearby large scale structures. If so, the Milky way and the Local
Group may be part of an extended filament in the cosmic web. However,
it may be noted that the location of these anisotropy spikes may have
some uncertainty due to their proximity to the zone of avoidance which
is artificially filled by mirroring galaxies from the unmasked
regions.

\citet{pandey15} applied an information theory based method
\citep{pandey13} to the SDSS Main galaxy sample and find that the
galaxy distribution in the Main sample is homogeneous on a scale of
$140 \hmpc$. The tests of isotropy in the present analysis is also
based on the information theory. Our analysis indicates that besides
the highly anisotropic nature of the present day galaxy distribution
on small scales, the Universe is isotropic around us on scales beyond
$90 \hmpc$. This reaffirms the validity of the assumption of
statistical isotropy on large scales and strengthens the foundations
of the standard cosmological model.

\section{ACKNOWLEDGEMENT}
The author thanks an anonymous reviewer for the valuable comments and
suggestions. The author thanks the 2MRS team for making the data
public. The author also thanks Rishi Khatri for his help in using
HEALPix. B.P. would like to acknowledge financial support from the
SERB, DST, Government of India through the project
EMR/2015/001037. B.P. would also like to acknowledge IUCAA, Pune and
CTS, IIT, Kharagpur for providing support through associateship and
visitors programme respectively.

\bsp	
\label{lastpage}

\begin{thebibliography}{99}

\bibitem[Akrami et al.(2014)]{akrami} Akrami, Y., Fantaye, Y.,
  Shafieloo, A., et al.\ 2014, \apjl, 784, L42

\bibitem[Alonso et al.(2015)]{alonso} Alonso, D., Salvador, A.~I.,
  S{\'a}nchez, F.~J., et al.\ 2015, \mnras, 449, 670

\bibitem[Appleby \& Shafieloo(2014)]{appleby} Appleby, S., \& Shafieloo, A.\ 2014, \jcap, 10, 070 

\bibitem[Barrow \& Hervik(2010)]{barrow} Barrow, J.~D.,
  \& Hervik, S.\ 2010, \prd, 81, 023513

\bibitem[Bengaly et al.(2015)]{bengaly} Bengaly, C.~A.~P., Jr., 
Bernui, A., \& Alcaniz, J.~S.\ 2015, \apj, 808, 39 

\bibitem[Bilicki et al.(2011)]{bilicki} Bilicki, M., Chodorowski, M.,
  Jarrett, T., \& Mamon, G.~A.\ 2011, \apj, 741, 31

\bibitem[Bilicki et al.(2014)]{bilicki1} Bilicki, M.,
  Jarrett, T.~H., Peacock, J.~A., Cluver, M.~E., \& Steward, L.\ 2014,
  \apjs, 210, 9

\bibitem[Blake \& Wall(2002)]{blake} Blake, C., \& Wall, J.\ 2002,
  \nat, 416, 150

\bibitem[Branchini et al.(2012)]{branchini} Branchini, E., Davis, M.,
  \& Nusser, A.\ 2012, \mnras, 424, 472

\bibitem[Briggs et al.(1996)]{briggs} Briggs, M.~S., Paciesas, W.~S.,
  Pendleton, G.~N., et al.\ 1996, \apj, 459, 40

\bibitem[Campanelli et al.(2011)]{campanelli} Campanelli, L., Cea, 
P., Fogli, G.~L., \& Marrone, A.\ 2011, \prd, 83, 103503 

\bibitem[\protect\citeauthoryear{Colles et al.}{2001}]{colles} 
Colles, M. et al.(for 2dFGRS team) 2001,\mnras,328,1039        

\bibitem[Dai et al.(2013)]{dai} Dai, L., Jeong, D., 
Kamionkowski, M., \& Chluba, J.\ 2013, \prd, 87, 123005 

\bibitem[Davis et al.(2011)]{davis} Davis, M., Nusser, A., Masters,
  K.~L., et al.\ 2011, \mnras, 413, 2906

\bibitem[Erdo{\v g}du et al.(2006b)]{erdogdu1} Erdo{\v g}du, P., Lahav,
  O., Huchra, J.~P., et al.\ 2006, \mnras, 373, 45

\bibitem[Erdo{\v g}du et al.(2006a)]{erdogdu2} Erdo{\v g}du, P.,
  Huchra, J.~P., Lahav, O., et al.\ 2006, \mnras, 368, 1515

\bibitem[Erdo{\v g}du \& Lahav(2009)]{erdogdu3} Erdo{\v g}du, P., \&
  Lahav, O.\ 2009, \prd, 80, 043005

\bibitem[Eriksen et al.(2007)]{eriksen} Eriksen, H.~K., Banday, A.~J.,
  G{\'o}rski, K.~M., Hansen, F.~K., \& Lilje, P.~B.\ 2007, \apjl, 660,
  L81
\bibitem[Fixsen et al.(1996)]{fixsen} Fixsen, D.~J., Cheng, E.~S.,
  Gales, J.~M., et al.\ 1996, \apj, 473, 576

\bibitem[G{\'o}rski et al.(1999)]{gorski1} Gorski, K.~M.,
  Wandelt, B.~D., Hansen, F.~K., Hivon, E., \& Banday, A.~J.\ 1999,
  arXiv:astro-ph/9905275

\bibitem[G{\'o}rski et al.(2005)]{gorski2} G{\'o}rski,
  K.~M., Hivon, E., Banday, A.~J., et al.\ 2005, \apj, 622, 759

\bibitem[Gruppuso et al.(2013)]{grupp} Gruppuso, A., Natoli, 
P., Paci, F., et al.\ 2013, \jcap, 7, 047 

\bibitem[Gupta \& Saini(2010)]{gupta} Gupta, S., \& Saini,
  T.~D.\ 2010, \mnras, 407, 651

\bibitem[Hanson \& Lewis(2009)]{hanlewis} Hanson, D., \&
  Lewis, A.\ 2009, \prd, 80, 063004

\bibitem[Hazra \& Shafieloo(2015)]{hazra} Hazra, D.~K., \& Shafieloo,
  A.\ 2015, \jcap, 11, 012

\bibitem[Hoftuft et al.(2009)]{hoftuft} Hoftuft, J., Eriksen, H.~K.,
  Banday, A.~J., et al.\ 2009, \apj, 699, 985

\bibitem[Huterer et al.(2015)]{huterer} Huterer, D., Shafer, D.~L., \&
  Schmidt, F.\ 2015, \jcap, 12, 033

\bibitem[Huchra et al.(2012)]{huchra} Huchra, J.~P., Macri, L.~M.,
  Masters, K.~L., et al.\ 2012, \apjs, 199, 26

\bibitem[Jackson(2012)]{jackson} Jackson, J.~C.\ 2012, \mnras, 426,
  779
     
\bibitem[Javanmardi et al.(2015)]{javanmardi} Javanmardi, B., 
Porciani, C., Kroupa, P., \& Pflamm-Altenburg, J.\ 2015, \apj, 810, 47 

\bibitem[Kaiser(1987)]{kaiser} Kaiser, N.\ 1987, \mnras, 227, 1

\bibitem[Kalus et al.(2013)]{kalus} Kalus, B., Schwarz, D.~J., Seikel,
  M., \& Wiegand, A.\ 2013, \aap, 553, A56

\bibitem[Kashlinsky et al.(2008)]{kashlinsky1} Kashlinsky, A., 
Atrio-Barandela, F., Kocevski, D., \& Ebeling, H.\ 2008, \apjl, 686, L49 

\bibitem[Kashlinsky et al.(2010)]{kashlinsky2} Kashlinsky, A.,
  Atrio-Barandela, F., Ebeling, H., Edge, A., \& Kocevski, D.\ 2010,
  \apjl, 712, L81

\bibitem[Kochanek et al.(2001)]{kochanek} Kochanek, C.~S., Pahre, M.~A., Falco, E.~E., et al.\ 2001, \apj, 560, 566 

\bibitem[Kogut et al.(1993)]{kogut} Kogut, A., Lineweaver, C., Smoot,
  G.~F., et al.\ 1993, \apj, 419, 1

\bibitem[Kov{\'a}cs \& Garc{\'{\i}}a-Bellido(2016)]{kovac} Kov{\'a
}cs, A., \& Garc{\'{\i}}a-Bellido, J.\ 2016, \mnras, 462, 1882

\bibitem[Klypin et al.(2003)]{klypin} Klypin, A., Hoffman, Y.,
  Kravtsov, A.~V., \& Gottl{\"o}ber, S.\ 2003, \apj, 596, 19

\bibitem[Land \& Magueijo(2005)]{land} Land, K., \& Magueijo, J\ 2005,
  \prl, 95, 071301

\bibitem[Lavaux et al.(2010)]{lavaux} Lavaux, G., Tully, R.~B.,
  Mohayaee, R., \& Colombi, S.\ 2010, \apj, 709, 483

\bibitem[Lin et al.(2016)]{lin} Lin, H.-N., Wang, S., Chang, Z., \&
  Li, X.\ 2016, \mnras, 456, 1881

\bibitem[Lynden-Bell et al.(1989)]{lyndenbell} Lynden-Bell, D., Lahav,
  O., \& Burstein, D.\ 1989, \mnras, 241, 325

\bibitem[Maller et al.(2003)]{maller} Maller, A.~H., McIntosh, D.~H.,
  Katz, N., \& Weinberg, M.~D.\ 2003, \apjl, 598, L1

\bibitem[Marinoni et al.(2012)]{marinoni} Marinoni, C., Bel, J., \&
  Buzzi, A.\ 2012, \jcap, 10, 036

\bibitem[Marozzi \& Uzan(2012)]{marrozi} Marozzi, G., \& Uzan,
  J.-P.\ 2012, \prd, 86, 063528

\bibitem[McCall(2014)]{mccall} McCall, M.~L.\ 2014, \mnras, 440, 405

\bibitem[Meegan et al.(1992)]{meegan} Meegan, C.~A., Fishman, 
G.~J., Wilson, R.~B., et al.\ 1992, \nat, 355, 143 

\bibitem[Moss et al.(2011)]{moss} Moss, A., Scott, D., Zibin, J.~P.,
  \& Battye, R.\ 2011, \prd, 84, 023014

\bibitem[Rowan-Robinson et al.(2000)]{rowan} Rowan-Robinson, M.,
  Sharpe, J., Oliver, S.~J., et al.\ 2000, \mnras, 314, 375

\bibitem[Pandey(2013)]{pandey13} Pandey, B.\ 2013, \mnras, 430, 
3376 
\bibitem[Pandey(2016)]{pandey16a} Pandey, B.\ 2016, \mnras, 462, 1630 

\bibitem[Pandey(2016)]{pandey16b} Pandey, B.\ 2016, \mnras, 463, 4239

\bibitem[Pandey \& Sarkar(2015)]{pandey15} Pandey, B., \& Sarkar,
  S.\ 2015, \mnras, 454, 2647

\bibitem[Peebles et al.(2001)]{peebles} Peebles, P.~J.~E.,
Phelps, S.~D., Shaya, E.~J., \& Tully, R.~B.\ 2001, \apj, 554, 104

\bibitem[Penzias \& Wilson(1965)]{penzias} Penzias, A.~A., \& Wilson,
  R.~W.\ 1965, \apj, 142, 419

\bibitem[Planck Collaboration et al.(2014b)]{aghanim} Planck
  Collaboration, Aghanim, N., Armitage-Caplan, C., et al.\ 2014, \aap,
  571, A27

\bibitem[Planck Collaboration et al.(2014a)]{adeplanck1} Planck
  Collaboration, Ade, P.~A.~R., Aghanim, N., et al.\ 2014, \aap, 571,
  A23

\bibitem[Planck Collaboration et al.(2016b)]{adeplanck2}
  Planck Collaboration, Ade, P.~A.~R., Aghanim, N., et al.\ 2016,
  \aap, 594, A16

 \bibitem[Planck Collaboration et al.(2016a)]{adeplanck3} Planck
   Collaboration, Ade, P.~A.~R., Aghanim, N., et al.\ 2016, \aap, 594,
   A13

\bibitem[Schwarz et al.(2004)]{schwarz1} Schwarz, D.~J., Starkman,
  G.~D., Huterer, D., \& Copi, C.~J.\ 2004, \prl,
  93, 221301

\bibitem[Schwarz \& Weinhorst(2007)]{schwarz2} Schwarz,
  D.~J., \& Weinhorst, B.\ 2007, \aap, 474, 717

\bibitem[Shannon(1948)]{shannon48} Shannon, C. E. \ 1948, Bell
System Technical Journal, 27, 379-423, 623-656

\bibitem[Scharf et al.(2000)]{scharf} Scharf, C.~A., Jahoda, K.,
  Treyer, M., et al.\ 2000, \apj, 544, 49

\bibitem[Smoot et al.(1992)]{smoot} Smoot, G.~F., Bennett, C.~L.,
  Kogut, A., et al.\ 1992, \apjl, 396, L1

\bibitem[Soda(2012)]{soda} Soda, J.\ 2012, Classical and 
Quantum Gravity, 29, 083001 

\bibitem[Tully et al.(2008)]{tully} Tully, R.~B., Shaya,
  E.~J., Karachentsev, I.~D., et al.\ 2008, \apj, 676, 184-205

\bibitem[Tully \& Fischer(1987)]{tully1} Tully, R.~B. \& Fischer,
  J.~R.\ 1987, Nearby Galaxies Atlas, Cambridge University Press

\bibitem[Watkins et al.(2009)]{watkins} Watkins, R., Feldman, H.~A.,
  \& Hudson, M.~J.\ 2009, \mnras, 392, 743

\bibitem[Wilson \& Penzias(1967)]{wilson} Wilson, R.~W., \& Penzias,
  A.~A.\ 1967, Science, 156, 1100

\bibitem[Wu et al.(1999)]{wu} Wu, K.~K.~S., Lahav, O., \& Rees,
  M.~J.\ 1999, \nat, 397, 225

\bibitem[Yahil et al.(1980)]{yahil} Yahil, A., Sandage, A., \&
  Tammann, G.~A.\ 1980, \apj, 242, 448

\bibitem[York et al.(2000)]{york} York, D.~G., et al.\ 2000, \aj,
  120, 1579

\bibitem[Zitrin \& Brosch(2008)]{zitrin} Zitrin, A., \& Brosch,
  N.\ 2008, \mnras, 390, 408

\end{thebibliography}
\end{document}